\documentclass[twocolumn]{aastex701}

\begin{document}

\title{Gravitomagnetic-Hydrodynamics and Turbulence in Early Universe}

\author[0009-0001-8646-9111]{Jia-Xiang Liang}
\affiliation{International Centre for Theoretical Physics Asia-Pacific, University of Chinese Academy of Sciences, Beijing 100190, P. R. China}
\affiliation{Center for Gravitational Wave Experiment, National Microgravity Laboratory, Institute of Mechanics, Chinese Academy of Sciences, Beijing 100190, China}
\email{liangjiaxiang25@mails.ucas.ac.cn}

\author[0000-0002-3543-7777]{Peng Xu}
\affiliation{Center for Gravitational Wave Experiment, National Microgravity Laboratory, Institute of Mechanics, Chinese Academy of Sciences, Beijing 100190, China}
\affiliation{Hangzhou Institute for Advanced Study, University of Chinese Academy of Sciences, Hangzhou 310024, China}
\affiliation{Taiji Laboratory for Gravitational Wave Universe (Beijing/Hangzhou), University of Chinese Academy of Sciences, Beijing 100049, China}
\email[show]{xupeng@imech.ac.cn}

\author[0000-0003-2155-3280]{Minghui Du}
\affiliation{Center for Gravitational Wave Experiment, National Microgravity Laboratory, Institute of Mechanics, Chinese Academy of Sciences, Beijing 100190, China}
\email{duminghui@imech.ac.cn}

\author[]{Yifu Cheng}
\affiliation{Key Laboratory for Mechanics in Fluid Solid Coupling Systems, Institute of Mechanics, Chinese Academy of Sciences, Beijing 100190, China}
\email{chengyifu@imech.ac.cn}

\author[0000-0003-4393-2118]{Zhan Wang}
\affiliation{Key Laboratory for Mechanics in Fluid Solid Coupling Systems, Institute of Mechanics, Chinese Academy of Sciences, Beijing 100190, China}
\email{zwang@imech.ac.cn}

\author[0000-0002-9533-8025]{Zi-Ren Luo}
\affiliation{Center for Gravitational Wave Experiment, National Microgravity Laboratory, Institute of Mechanics, Chinese Academy of Sciences, Beijing 100190, China}
\affiliation{Hangzhou Institute for Advanced Study, University of Chinese Academy of Sciences, Hangzhou 310024, China}
\affiliation{Taiji Laboratory for Gravitational Wave Universe (Beijing/Hangzhou), University of Chinese Academy of Sciences, Beijing 100049, China}
\email{luoziren@imech.ac.cn}

\author[]{Man-Jia Liang}
\affiliation{Center for Gravitational Wave Experiment, National Microgravity Laboratory, Institute of Mechanics, Chinese Academy of Sciences, Beijing 100190, China}
\email{liangmanjia21@mails.ucas.ac.cn}

\begin{abstract}


The nonlinear coupling between spacetime geometry and matter in the early Universe remains a frontier in theoretical cosmology. By introducing a novel gravitomagnetic-hydrodynamic framework, we reveal a fundamental analogy between magnetohydrodynamics and the co-evolution of spacetime geometry and relativistic plasma. We demonstrate that, in high-energy environments such as the electroweak phase transition, the (newly defined) gravitomagnetic Reynolds number becomes large, signifying a strongly coupled system where the gravitomagnetic field could be frozen into the fluid. This coupling inevitably leads to the emergence of gravitational Alfvén waves and could drive a transition to turbulence involving the dynamics of spacetime itself. Our findings suggest that gravitomagnetic-hydrodynamic turbulence may leave imprints on the stochastic gravitational wave background, offering a new window into the nonlinear dynamics of the primordial Universe.

\end{abstract}




\section{Introduction}\label{sec1}


The fluid model can be viewed as the most successful model for macroscopic and continuous matter systems, that has long underpinned diverse fields in physics and engineering. 
Within such a framework, by statistical averaging of microscopic interactions, essential properties from the microscopic world brought up to the coarse-grained continuum descriptions are encoded into the macroscopic observables and their relations (e.g., equations of state and constitutive relations). 
Integrated with conservation laws, the fluid dynamics provides a universal language for characterizing the macroscopic behaviors of complex matter systems across from laboratory to cosmological scales with a finite number of such observables or measurable quantities, see Landau's seminal treatise \citep{landau1987fluid}.

Based on Einstein's revolutionary vision of nature as geometry, space, time, and matter are bound together in a formalism called geometrodynamics according to his general theory of relativity (GR) \citep{einstein1915erklarung,misner1973gravitation}.
Today, confronted with the many stringent tests since the last century \citep{will2014confrontation,will1953theory}, GR stands still as the best-fit theory of gravitation or dynamical spacetime among the many alternatives, and serves as another universal language for understanding a wide range of physical processes including the dynamics of spacetime itself. 
Following the studies of black hole thermodynamics, it was suggested in \citep{jacobson1995thermodynamics} that geometrodynamics, just like macroscopic fluid dynamics, might be a coarse-grained dynamics of collective variables from the underlying but still unknown quantum geometry. 
The gravity-fluid analogy \citep{andersson2021relativistic} also provides us with clues about the deep connection between these two ``universal models'' of the macroscopic world. 
Recently, new clues as turbulence in dynamical spacetime have been studied.
It was shown that black hole horizons exhibit statistically steady spacetime turbulence, consistent with Kolmogorov’s theory \citep{andrade2021driven,yang2015turbulent}.
A theory for four-wave interactions of gravitational waves was also established, and both direct and inverse cascades were found for the energy and wave action respectively \citep{galtier2017turbulence}.  

On the other hand, the search for a fully self-consistent theory of relativistic fluid mechanics remains an unfinished journey, see \citep{andersson2021relativistic} for a comprehensive review. 
When heat flows, strong nonlinearities, and dissipation are considered, the formalism of relativistic fluid mechanics turns out to be complicated \citep{carter1985canonical} and brings debates and objections in the literature \citep{garcia2009nature}, which also lead to insights into the deep connections among causality, stability, and hyperbolicity in the fully-fledged relativistic fluid model \citep{hiscock1983stability}.
While, such properties are essential in modeling high-energy and complex hydrodynamical systems in astrophysics and cosmology. 
See, for example, \citep{mizuno2006grmhd,narayan2012grmhd,fernandez2019long,moscibrodzka2009radiative,cattorini2024grmhd,liska2020large} for recent developments in the study of complex Magneto-HydroDynamics (MHD) under strong gravitation.

Attempts of new methods and investigations in relativistic fluid models could help deepen our understanding of the highly nonlinear and complex geometrodynamics, and shed new light on the studies of the physical processes in the early Universe and high-energy astrophysics. 
In this work, we attempt to follow a different line of thought.
In the weak field limit, due to the preservation (approximately) of the Lorentz symmetry, Einstein's equation could be cast into equations similar to Maxwell's equations of electromagnetism (EM) \citep{mashhoon2003gravitoelectromagnetism}.
Such formalism, known as Gravito-electromagnetism (GEM), and its rich correspondences with electromagnetism were gradually established for the full theory of GR \citep{bel1958cr,penrose1960spinor,maartens1998gravito,campbell1971debye,misner1973gravitation,nichols2011visualizing,nichols2012visualizing,zhang2012visualizing}.  
One key feature is that the gravitomagnetic (GM) part of the spacetime curvature will couple to the matter currents in a similar way to the couplings between magnetic fields and electric currents.

Such an analogy between GEM and EM had already proven very useful and had found applications in understanding the strong interactions between fluids and gravity in the studies of active galactic nuclei and their jets \citep{thorne1986black}. With the extended Navier-Stokes equations \citep{ciufolini1995gravitation}, the hydrodynamics of an accretion disk driven also by the GM field could be studied.
Meanwhile, MHD theory reveals the bounded dynamics of the fluid and the solenoidal magnetic field \citep{galtier2016introduction}. We take one step further along this line and suggest a new counterpart in the GEM landscape analogous to electromagnetism \citep{liang2024gravitomagnetohydrodynamicsspacetimeturbulenceearly}, a counterpart GravitoMagnetic-Hydrodynamics (GMHD) model that could help us understand the possible new physics of highly bounded and dissipative spacetime-matter systems.





\section{Gravitomagnetic-Hydrodynamics in early universe}

In our Universe, most of the baryonic matter is in the form of a plasma state. 
In the 1940s, Hannes Alfvén established the basic theory and framework of MHD for plasmas by considering the combination of Maxwell's equations with the Navier–Stokes equation in hydrodynamics \citep{alfven1942existence}. 
In this framework, plasmas can be described by a specific fluid model, that a conducting fluid interacting with magnetic fields. 
Generally, MHD fluids in the Universe are far from thermodynamic equilibrium and may exhibit highly turbulent dynamics (see \citep{galtier2016introduction} for details).


In the cosmological context, MHD is not merely governed by magnetic field interactions but is also subject to the fundamental laws of gravity or geometrodynamics.
In GR, given a 3+1 foliation of spacetime with respect to a congruence of observers with 4-velocity $z^{\mu}$.
The Weyl curvature tensor can be split into two irreducible parts, the electric and magnetic parts, defined in the space slice with respect to $z^{\mu}$ \citep{nichols2011visualizing},
which can be viewed as mimicking the separation of the Maxwell tensor $F_{\mu\nu}$ into electric and magnetic fields. Such analog extends to the field equations \citep{ciufolini1995gravitation,costa2021frame}, that reference to the free-falling observers the Einstein's equation can be expanded as
\begin{eqnarray}
    \nabla\cdot E_{g}&=&4\pi G\rho+ \mathcal{O}(E_g^2)+\mathcal{O}(B_g^2)+...,\label{eq:Eg}\\ 
    \nabla \times E_{g}&=&-\frac{1}{2c}\frac{\partial}{\partial t}B_{g}+\mathcal{O}(E_g\times B_g)+...,\label{eq5}\\
    \nabla\cdot B_{g}&=&0+\mathcal{O}(E_g\cdot B_g)+...,\label{eq:Bg}\\
    \nabla\times B_{g}&=&\frac{8\pi G}{c}J_{g}+\frac{2}{c}\frac{\partial}{\partial t}E_{g}+\mathcal{O}(E_g\times B_g)+... \ \label{eq4}   .
\end{eqnarray}
Geodesic motion of a test mass is equivalent to a Lorentz-like acceleration, 
\begin{equation}
    a=-E_{g}-\frac{u}{2c}\times B_{g} + ...\:.\label{eq:Lorentz}
\end{equation}
Here, $E_g$ and $B_g$ denote the gravitoelectric (GE) and gravitomagnetic fields, and $\rho$ and $J_g$ the matter density and current, and $u$ the velocity. 
The linear parts of the above equations just resemble the Maxwell-like equations, and complicated nonlinear coupling terms appear when approaching the strong field regime, similar to the case of nonlinear electrodynamics. Even for full GR, such analogues retain still in some sense, see \citep{nichols2011visualizing} for details.  In this work, we aim to investigate how such correspondence may deepen our understanding of the co-evolution of spacetime geometry and relativistic fluid, particularly in regimes where the characteristic fluid velocity $u$ approaches the speed of light $c$ as in the case of the early Universe.  Therefore, as the preliminary step along this line \citep{liang2024gravitomagnetohydrodynamicsspacetimeturbulenceearly}, this work will focus on the leading-order parts of the above GEM equations and establish the essential characteristics of GMHD fluids. The full non-linear geometrodynamics cast in the GEM form will be addressed in future studies.

Back to the MHD case,  the generalized Ohm's law, incorporating dissipation effects, plays an important role in simplifying and reformulating the MHD equations \citep{galtier2017turbulence}.
In electromagnetism, Ohm's law establishes a linear proportionality between the driven electric field and the current density, with dissipation arising from random scatterings of current carriers due to disordered collisions. This scattering impedes electric field propagation and dissipates energy \citep{di2004disordered}.
Following the same line, for relativistic fluids in the early Universe where gravitational interactions dominate, the complicated high frequency or small scale interactions can be coarse-grained, and dissipative effects at large (cosmic) scales can be manifested as the ``effective'' gravito-conductivity $\sigma_g$,
\begin{equation}
    J_{g}=\sigma_{g}(E_{g}+\frac{u}{2c}\times B_{g}) \label{eq:Ohm}.
\end{equation}
Therefore, a generalized gravitational Ohm's law, that the constitutive relation between the matter current $J_g=\rho u$ and the GE field $E_g$ (generated by the energy density) can be established.  The second term,  similar to a motional electromotive force, is from the Lorentz force-like term in Eq. (\ref{eq:Lorentz}).
The combination of gravitational Ohm's law and the GEM field equations (\ref{eq5}) and (\ref{eq4}) leads us to one of the basic equations of GMHD
\begin{equation}
     \frac{\partial}{\partial t}B_{g}=\nabla\times(u\times B_{g})-\eta_g\nabla\times(\nabla\times B_{g}-\frac{2}{c}\frac{\partial}{\partial t}E_{g})\:. \label{eq:GMHD-B}
\end{equation}
Here, we define the GM diffusivity as $\eta_{g}=\frac{c^2}{4\sigma_{g}\pi G}$, similar to the magnetic diffusivity $\eta_{m}=\frac{1}{4\sigma\pi \epsilon_{0}}$, which provides the transport coefficient under gravitational and other dissipative effects. Combining Eq.(\ref{eq:GMHD-B}) with the energy-momentum conservation under dissipation and the remaining GEM field equations (Eq. (\ref{eq:Eg}) and (\ref{eq:Bg})) yields the fundamental GMHD equations governing the coupled dynamics of relativistic fluids and spacetime geometry. The GMHD equations assume different forms for various physical situations. In the Appendix (\ref{secA1}), we present the specific case of an ideal, incompressible GMHD system derived in the vanishing-dissipation limit.
Although the full nonlinear geometrodynamics is absent in this work, the first-order analysis of GMHD fluids has already revealed interesting physical pictures and the co-evolution of the bounded spacetime-matter system.

According to  Kelvin’s Theorem, for any solenoidal field $F$ (like the GM field in Eq. (\ref{eq:Bg}) ) and a material surface $S$ moving with the fluid, we have 
\begin{equation}
    \frac{d}{d t}\int_S F\cdot dS=\int_S\left(\frac{\partial}{\partial t}F- \nabla\times(u\times F)\right)\cdot dS\;.
\end{equation}
Applied to GMHD, in the limit of an ideal GMHD fluid ($\sigma_g\to \infty, \eta_g\to 0$), the GM flux through any material surface advected by the fluid is conserved,
 \begin{equation}
     \frac{d}{d t}\int_S B_{g}\cdot dS=0\;.
 \end{equation}
This is consistent with Alfvén's Theorem. It means that the GM flux through any closed surface $S$ moving with the fluid is conserved. In other words, the GM field is frozen into the GMHD fluid in the sense that it moves with the fluid, which can be viewed as a significant coupling effect of relativistic fluids and GM fields. 
In MHD, we use the magnetic Reynolds number $R_{m}=\frac{uL}{\eta_m}$ to  measure the strength of this coupling, where L denotes the scale of the fluid system under consideration. We introduce a corresponding GM Reynolds number $R_{g}=\frac{uL}{\eta_{g}}$. Similar to the MHD theory, in the limit $R_{g}\to \infty$ (equivalently $\eta_g\to 0,\sigma_g \to \infty$) the GMHD fluid becomes effectively ideal. Thus, the dynamics of the fluid and the GM field become tightly coupled, and the generalized Alfvén wave in GMHD systems will naturally emerge.

The key condition for the validity of the above physical picture of strong coupling is whether there exist possible situations in which the GMHD system could be sufficiently close to ideal. In the following, we provide an estimation of $\sigma_{g},\eta_{g}$ for relativistic fluids in the early Universe and prove the existence of strongly coupled GMHD systems.


For plasmas in the early universe, the kinematic viscosity $\nu=\frac{\eta}{\rho+p}$ \citep{caprini2009stochastic} and $\eta$ is the shear viscosity. 
Given the high temperature of primordial plasma, the shear viscosity has the form \citep{arnold2000transport}
\begin{equation}
    \eta=C_1\frac{T^3}{\chi^4 \ln \chi^{-1}}\;,
\end{equation}
where $T$ is the temperature, $\chi$ the gauge coupling, and $C_1$ a constant. 
The kinematic viscosity is the transport coefficient that characterizes the diffusion of transverse momentum due to collisions in a medium. The most important viscosity comes from the weakest interactions, since it is inversely proportional to the scattering cross-section of the related processes. Before the electroweak phase transition, viscosity is dominated by right-handed lepton transport, while after the transition, it is dominated by neutrinos \citep{caprini2009stochastic},
\begin{equation}\label{eq13}
    \nu(T)\approx\left\{
    \begin{array}{ll}
    22T^{-1} & T>100\ \mathrm{GeV} \\
    5\times 10^{8}\ \mathrm{GeV}^{4}T^{-5} & 100~\mathrm{MeV} \leq T \leq 100~\mathrm{GeV} \\
    2\times 10^{9}\ \mathrm{MeV}^{4}T^{-5} & T<100\ \mathrm{MeV}
    \end{array}
    \right.
\end{equation}
According to the estimates in \citep{giovannini2012reynolds}, the Reynolds number of the fluid in the early universe is about $R=10^{16}$. This shows that the fluid in the early universe is close to an ideal fluid.


An estimate for the gravitomagnetic Reynolds number $R_g$ (or equivalently, the effective gravito-conductivity $\sigma_g$) can be obtained from the generalized Ohm’s law in Eq. (\ref{eq:Ohm}) and the order-of-magnitude estimations of fluctuations in the primordial fluid through the associated stochastic gravitational wave background (SGWB). 

To confront with observations, the energy density $\rho_{GW}$ of the SGWB, produced by disturbances like first-order phase transitions, etc.,  can be defined as 
\begin{equation}
    \rho_{GW}=\Omega_{GW}\rho_{c} ,
\end{equation}
where $\rho_{c}$ is the critical energy density and $\Omega_{GW}$ is the ratio of gravitational wave energy density to the critical density of the flat Universe. 
 In the GEM formalism, $\rho_{GW}$ can be written down in terms of $E_{g}$ and $B_{g}$ \citep{nichols2011visualizing}
\begin{equation}
    \rho_{GW} = \frac{|E_{g}|^{2} + \frac{1}{4} |B_{g}|^{2}}{8 \pi G}=\frac{|B_{g}|^{2}}{16 \pi  G}\;.
\end{equation}
The second equality is due to the equipartition of energy among different degrees of freedom for SGWB. 
As a radiation field, gravitational wave energy density scales with the scale factor $a$ as
\begin{equation}
    \rho_{G W }^*=\rho_{G W}^0\left(\frac{a^0}{a^*}\right)^4\;,
\end{equation}
and 
\begin{equation}
    \frac{a^0}{a^*}=\frac{T^*}{T^0}\left(\frac{g^{*}}{g^{0}}\right)^{1/3}\;.
\end{equation}
Here the indices * and 0 stand for the time of primordial SGWB generation (e.g. during the electroweak phase transition) and the present time, respectively, and $g$  is the number of relativistic degrees of freedom.
Therefore, given the present day observations, the magnitude of the GM field in the early Universe can be derived as 
\begin{equation}
    |B_{g}|^*=4\sqrt{\pi G\rho_{G W }^*}=4\sqrt{\pi G\rho_{G W}^0}\left(\frac{T^*}{T^0}\right)^{2}\left(\frac{g^{*}}{g^{0}}\right)^{2/3}\;.
\end{equation}
According to the generalized Ohm's law in Eq. (\ref{eq:Ohm}), also considering the equipartition of energy, we then have the relation 
\begin{equation}
   \alpha |J_{g}|^*= \sigma_{g}^*|B_{g}|^*\;,
\end{equation}
where $\alpha\sim 10^{-1}$ is a dimensionless constant introduced to parameterize the approximation errors.
$\sigma_{g}$ is estimated via Eq. (\ref{eq:Ohm}), with the fluid density postulated to be comparable to the cosmological critical density $\rho$ at that epoch. 

For the relativistic fluids in the early Universe, it is natural to assume that its energy density is comparable to the critical mass density at that epoch
\begin{eqnarray}
    \rho^*\approx\frac{3H^{2}}{8\pi G},\ \ \ 
    H=c^{-\frac{5}{2}}G^{\frac{1}{2}}\hbar^{-\frac{3}{2}}k_B^{2}(T^{*})^{2}\; .
\end{eqnarray}
Given the bubble wall velocity  $|v_{w}|\sim0.87c$ in phase transitions in the early Universe, the velocity of the relativistic fluid should approach the speed of light. 
 Therefore, the gravito-conductivity reads 
\begin{equation}
    \sigma_g^* = \alpha\frac{3}{32\pi} \frac{k_B^4 (T^0T^*)^2}{c^4 \hbar^3 \sqrt{\pi G \rho_{GW}^0}} \left( \frac{g^0}{g^*} \right)^{2/3}\; ,\label{eq:sigmag}
\end{equation}
and the GM diffusivity can be calculated as
\begin{equation}
    \eta_g^*= \frac{8}{3} \frac{c^6 \hbar^3 \sqrt{\pi \rho_{GW}^0}}{\alpha k_B^4 (T^0 T^*)^2 \sqrt{G}} \left( \frac{g^*}{g^0} \right)^{2/3}\;.
\end{equation}
Remarkably, the gravito-conductivity exhibits a temperature dependence $\sigma_g=C_2\frac{T^{2}}{\sqrt{G}}$, which mirrors, despite the profoundly different physical origins, the scaling of shear viscosity in high-temperature plasmas derived from Kubo’s formula for gauge theories $\eta=C_1\frac{T^3}{\chi^4 \ln \chi^{-1}}$. 
This unexpected parallel between a gravitational transport coefficient and a purely quantum-field-theoretic quantity underscores a deep and surprising structural analogy between GHMD and non-Abelian plasma dynamics.
Given the current constraint on $\Omega_{GW}^0\sim10^{-12}$ \citep{peter2018gravitational} and the suggested values for other parameters in Eq. (\ref{eq:sigmag}), the gravito-conductivity turns out to be $\sigma_g^*\sim10^{-9} kg\:s\:m^{-3}\:K ^{-2}\:(T^*)^{2}$ and $\eta^*_{g}\sim\frac{ 10^{34}}{T_*^2}m^2\:s^{-1}\:K^{2}$ in the early Universe. This confirms our assumption made in the first half of this section that the gravito-conductivity attains a large value for the relativistic fluid in the extremely hot, dense early Universe, a regime where gravitational and thermal energies are intrinsically intertwined. Then, the GMHD fluids in this epoch are expected to be nearly ideal, with the fluids and GM fields tightly coupled. 
The existence of this GMHD picture in the early Universe is further reinforced by an explicit estimate of the GM Reynolds number $R_g$ for the concrete scenario of a first-order phase transition in the following section. 
The resulting $R_g\gg1$ strongly supports the validity of the ideal GMHD framework. 
With high Reynolds numbers, previous studies \citep{arnold2000transport,gogoberidze2007spectrum,dolgov2002relic,auclair2022generation,kosowsky2002gravitational} show that MHD turbulence exists after the electroweak phase transition. Meanwhile, due to a large $R_{g}$ or $\sigma_g$, it is then expected that GMHD turbulence will appear in a similar manner as for the case of MHD turbulence.

\begin{table*}
\centering
\caption{Parameters of  different hydrodynamic systems.}
\begin{tabular}{|c|c|c|c|}
\hline
\multicolumn{1}{|c|}{Dissipative system} & \multicolumn{1}{c|}{Transport coefficients} & \multicolumn{1}{c|}{Diffusion coefficients} & \multicolumn{1}{c|}{Reynolds number} \\
\hline
fluid & $\eta\ [Pa\cdot s]$& $\nu\ [m^2/s]$& $R$ \\
\hline
MHD & $\sigma\ [	
S\cdot m^{-1}]$& $\eta_{m}\ [m^2/s]$& $R_{m}$ \\
\hline
GMHD & $\sigma_{g}\ [kg\cdot s\cdot m^{-3}]$& $\eta_{g}\ [m^2/s]$& $R_{g}$ \\
\hline
\end{tabular}\label{table1}
\end{table*}

\section{GMHD Turbulence in early Universe}

In the early universe, there were a series of cosmological sources that produced violent disturbances, including first-order phase transitions and interactions between topological defects. These disturbances will produce fully developed turbulence in the plasma\cite{peter2018gravitational}. GMHD  turbulence can develop when the GM Reynolds number is sufficiently large.



The GM Reynolds number $R_{g}=\frac{uL}{\eta_{g}}$ is a parameter that determines the degree of GMHD turbulence nonlinearity.
Turbulence occurs when a large Reynolds number fluid is perturbed. For the first-order phase transition, the characteristic parameters of the fluid under consideration can be derived from the characteristic length scale of phase bubbles and the bubble wall velocity $v_{w}$. The collisions of bubbles strongly perturb the fluid, so the velocity and characteristic size of the fluid are comparable to the wall velocity and bubble size.
 According to the relationship between wall velocity $v_{w}$ and sound velocity in the fluid, there are three models: Subsonic deflagrations, Detonations, and Supersonic deflagrations (hybrids). In general, one has $v_{w}\approx 0.87c$ for the detonation case  and $v_{w}\approx c_{s}$  for deflagrations \citep{caprini2009stochastic} where $c_{s}=1/\sqrt{3}c$ for the sound velocity in relativistic fluids. 
It is natural to set the diameter of phase bubbles as the featured size of the fluid $L$ \citep{hindmarsh2021phase}.
The bubble size $L$ is comparable to the Hubble length, while the dimensionless  parameter $LH$ attains a magnitude of order $1\sim0.1$. Therefore, the GM Reynolds number is 
 \begin{equation}
     R_{g}=\frac{uL}{\eta_{g}}\sim\frac{3}{8}\alpha (T^0)^2 k_B^2 \left(\frac{g^0}{g^{*}}\right)^{2/3} \frac{(c\hbar)^{-3/2}}{\sqrt{\pi\rho^0_{GW}}}\;.
 \end{equation}
The GM Reynolds number exhibits an interesting correlation with the present-day gravitational wave energy density $\Omega_{GW}^0$. Given the present-day constraints $\Omega_{GW}^0=10^{-11}\sim10^{-15}$, the GM Reynolds number $R_{g}$ attains a magnitude of order $10^{3}\sim10^{5}$. Consequently, we propose that a first-order phase transition at sufficiently high energy scales can give rise to GMHD turbulence under the specified physical conditions and parameters. 

In the spirit of Alfvén’s foundational picture of MHD, Alfvén waves arise when electromagnetic disturbances propagate through an electrically conducting fluid. By direct analogy, gravitational Alfvén waves can emerge in the early Universe when spacetime perturbations couple to a relativistic fluid with a sufficiently large GM Reynolds number $R_g$. In this regime, the fluid velocity and GM field undergo sustained, coupled oscillations, manifesting as propagating gravitational Alfvén waves.  
A deeper perspective suggests that, through collisions of Alfvén wave packets, an energy cascade could be established in GMHD, channeling energy from large to small scales, ultimately dissipating as heat at the smallest (dissipative) scales.  
Based on dimensional analysis, isotropic GMHD turbulence is expected to exhibit an energy spectrum scaling as  $k^{-3/2}$,  a result analogous to the Iroshnikov–Kraichnan (IK) spectrum in MHD turbulence \citep{galtier2016introduction}.
A detailed derivation of this physical picture for the simple ideal and incompressible GMHD limit is provided in the Appendix (\ref{secA1}).
Crucially, this turbulent cascade involves both the relativistic fluid and the GM field on an equal footing. Consequently, the very geometrodynamics of spacetime itself, through its dynamical coupling to matter, can acquire turbulent characteristics. This striking notion aligns with recent theoretical developments suggesting that general relativistic spacetime itself may support turbulence-like behavior, as summarized in \citep{galtier2022physics,galtier2020plausible}.

Furthermore, we consider the leading-order continuity equation $\frac{\partial}{\partial t}\rho\approx-\nabla\cdot(\rho u)$.
On sufficiently small scales where $\sigma_{g}$ is spatially homogeneous, the continuity equation reduces to
\begin{equation}\label{eq16}
    \frac{\partial}{\partial t}\rho\approx -\sigma_{g}\nabla\cdot(E_{g}+\frac{u}{2c}\times B_{g})\;.
\end{equation}
Through dimensional analysis, the equation reduces to a form
\begin{equation}
    \frac{\partial}{\partial t}\rho\sim -C_{3}\sigma_{g} G \rho\;,
\end{equation}
where $C_{3}$ is a dimensionless constant. The solution can be written down as
\begin{equation}
    \rho(t) \sim \rho_0 e^{-\sigma_g C_3 G t}\;,
\end{equation}
which pointed out that the absolute value $|\sigma_{g}|$ scales the rate of density variation. Therefore, the onset of GMHD turbulence may amplify density perturbations, enhancing the GM Reynolds number $R_{g}$ and driving the system into a strongly nonlinear regime. This may produce an inverse cascade in GMHD turbulence. In \citep{galtier2020plausible}, an inverse cascade of wave action in gravitational wave turbulence leads to the expansion of the Universe. This is very similar to our analysis. 


\section{Concluding remarks}\label{sec13}

In this work, we have established the foundations of GMHD, a new theoretical framework that extends the formalism of magnetohydrodynamics to incorporate GEM fields in the weak-gravity regime. By deriving the governing equations and introducing key parameters such as gravito-conductivity $\sigma_g$, GM diffusivity $\eta_g$, and the GM Reynolds number $R_g$ (Table (\ref{table1}) lists the parameters for different hydrodynamic systems), we have demonstrated how spacetime and matter can become tightly coupled in high-energy environments.

Our estimates indicate that in the early Universe, particularly during epochs such as the electroweak phase transition, the GM Reynolds number reaches values sufficiently high to support fully developed turbulence. This GMHD turbulence is expected to exhibit an Iroshnikov–Kraichnan-type energy spectrum and may facilitate both direct and inverse cascades, potentially amplifying density perturbations and influencing the expansion history of the Universe.

Looking forward, the GMHD framework opens several promising avenues for future research. These include the study of full nonlinear GEM effects, the role of GMHD turbulence in phase transitions and inflation, and the search for observational signatures in the SGWB. By bridging relativistic fluid dynamics, gravitation, and turbulence theory, GMHD provides a powerful new paradigm for exploring the complex interplay between spacetime and matter in the early Universe.

\begin{acknowledgments}
This work is supported by the National Key Research, Development Program of China, Grant No. 2021YFC2201901, and the International Partnership Program of the Chinese Academy of Sciences, Grant No. 025GJHZ2023106GC.

\end{acknowledgments}

\appendix

\section{Alfvén speed and IK  energy spectrum }\label{secA1}
We model the early universe using a simplified GMHD framework. Given that the characteristic timescales are much shorter than the Hubble time, we neglect cosmic expansion. The cosmological principle implies homogeneous and isotropic energy density distribution. Since fluctuations in $E_{g}$ are subdominant to those in  $B_{g}$, we focus exclusively on  $B_{g}$ fluctuations.
Therefor, we neglect the $\frac{\partial}{\partial t}E_{g}$ in Eq. (\ref{eq4}), yielding:
\begin{equation}
    J_{g}=\frac{c}{8\pi G} \nabla\times B_{g}\;.
\end{equation}
For an ideal GMHD,  the  momentum conservation equation and Eq. (\ref{eq:GMHD-B}) reduces to
\begin{equation}
        \rho(\frac{\partial}{\partial t}u+u\cdot \nabla u)= \rho E_{g}+ J_{g}\times \frac{1}{2c}B_{g}-\nabla p\;, \\
    \frac{\partial}{\partial t}B_{g}=\nabla\times(u\times B_{g})\;.
\end{equation}
Consider the first order perturbation of $u$ and $B_{g}$. for $u=u_{1}, B_{g}=B_{0}+B_{1}$, and ignore higher order terms.
\begin{equation}
   \frac{\partial}{\partial t}u_{1}=\frac{1}{16\pi G\rho} (\nabla\times B_{1})\times B_{0}\approx\frac{1}{16\pi G\rho}(B_{0}\cdot\nabla)B_{1}\;,\\
\frac{\partial}{\partial t}B_{1}=\nabla\times(u_{1}\times B_{0})=(B_{0}\cdot\nabla)u_{1}\;.
\end{equation}
Combining these yields a wave equation:
\begin{equation}
    \frac{\partial^{2}}{\partial t^{2}}u_{1}=\frac{1}{16\pi G\rho}(B_{0}\cdot\nabla)\frac{\partial}{\partial t}B_{1}=\frac{1}{16\pi G\rho}(B_{0}\cdot\nabla)^{2}u_{1}\;.
\end{equation}
 In order to find the dispersion relation we
introduce
 \begin{equation}
     \hat{u}=u_{1}e^{i(kx-\omega t)},\   \   \   \   \hat{B_{g}}=B_{1}e^{i(kx-\omega t)}\;,
 \end{equation}
where the symbol  $\hat{}$ means the Fourier transform. For $\vec{B}=\left|B\right|\cdot e_{b}$, $\vec{k}\cdot\vec{B}=\left|B\right|\cdot k_{b}$, we derive the dispersion relation:
\begin{equation}
    \omega^{2} \hat{u}=\frac{1}{16\pi G\rho}(B_{0}\cdot k)^{2}\hat{u}\  \  
\  \  \  \omega^{2}=\frac{(B_{0}\cdot k)^{2}}{16\pi G\rho}\;,
\end{equation}
defining the gravitational Alfvén speed:$V_{B}=\frac{B_{0}}{4\sqrt{\pi G \rho}}$.

The IK phenomenology \citep{iroshnikov1964turbulence,kraichnan1965inertial} assumes that even if there is no uniform magnetic field, the turbulence structures mainly experience a locally uniform
magnetic field. In wave turbulence, energy is transferred through wave packets collisions processes. To simplify, We assume that wave packet  are uniform in different directions $z^{+}\sim z^{-}\sim z$($z^{\pm}$ is Elsasser variables, $z^{\pm}=u\pm V_{B}$). Then the energy transfer  satisfies, 
\begin{equation}\label{eq:tautransfer}
    \varepsilon\sim\frac{z_{l}^{2}}{\tau}\;.
\end{equation}
We labeled Elsasser variables $z_{l}$ with $l$ to indicate the scales.
 The turbulence energy transfer was attributed to wave packets collisions and the collision is pretty short, the contribution of once collision should be proportional to the $\tau_{A}$. With the help of dimensional analysis,
\begin{equation}
   \tau_{A}=\frac{l}{V_{B}}\;,  \  \  \ \Delta z_{l}\sim\tau_{\rm A}\frac{z^{2}_{l}}{l}\;.
\end{equation}
A random-walk model gives the energy transfer rate \citep{galtier2016introduction}:
\begin{equation}
\varepsilon \sim \frac{z_l^2}{\tau} \sim \frac{z_l^4}{l V_B}, 
\end{equation}
where \(z_l\) is the Elsässer variable amplitude at scale \(l\). Rewriting in terms of the energy spectrum \(E(k) \sim z_l^2 / k\), we find:
\begin{equation}
\varepsilon \sim \frac{E(k)^2 k^3}{V_B}, 
\end{equation}
leading to the isotropic IK spectrum for GMHD turbulence:
\begin{equation}
E(k) = C \sqrt{\varepsilon V_B} \, k^{-3/2}. 
\end{equation}
This  establishes the \(-3/2\) scaling for gravitational turbulence, analogous to magnetized plasmas.

\bibliography{sn-bibliography}

\begin{thebibliography}{}
\expandafter\ifx\csname natexlab\endcsname\relax\def\natexlab#1{#1}\fi
\providecommand{\url}[1]{\href{#1}{#1}}
\providecommand{\dodoi}[1]{doi:~\href{http://doi.org/#1}{\nolinkurl{#1}}}
\providecommand{\doeprint}[1]{\href{http://ascl.net/#1}{\nolinkurl{http://ascl.net/#1}}}
\providecommand{\doarXiv}[1]{\href{https://arxiv.org/abs/#1}{\nolinkurl{https://arxiv.org/abs/#1}}}

\bibitem[{H. Alfv{\'e}n(1942)Alfv{\'e}n}]{alfven1942existence}
Alfv{\'e}n, H. 1942, \bibinfo{title}{Existence of electromagnetic-hydrodynamic waves,} Nature, 150, 405

\bibitem[{N. Andersson \& G.~L. Comer(2021)Andersson \& Comer}]{andersson2021relativistic}
Andersson, N., \& Comer, G.~L. 2021, \bibinfo{title}{Relativistic fluid dynamics: Physics for many different scales,} Living reviews in relativity, 24, 3

\bibitem[{T. Andrade {et~al.}(2021)Andrade, Pantelidou, Sonner, \& Withers}]{andrade2021driven}
Andrade, T., Pantelidou, C., Sonner, J., \& Withers, B. 2021, \bibinfo{title}{Driven black holes: from Kolmogorov scaling to turbulent wakes,} Journal of High Energy Physics, 2021, 1

\bibitem[{P. Arnold {et~al.}(2000)Arnold, Moore, \& Yaffe}]{arnold2000transport}
Arnold, P., Moore, G.~D., \& Yaffe, L.~G. 2000, \bibinfo{title}{Transport coefficients in high temperature gauge theories (I): leading-log results,} Journal of High Energy Physics, 2000, 001

\bibitem[{P. Auclair {et~al.}(2022)Auclair, Caprini, Cutting, Hindmarsh, Rummukainen, Steer, \& Weir}]{auclair2022generation}
Auclair, P., Caprini, C., Cutting, D., {et~al.} 2022, \bibinfo{title}{Generation of gravitational waves from freely decaying turbulence,} Journal of Cosmology and Astroparticle Physics, 2022, 029

\bibitem[{L. Bel(1958)Bel}]{bel1958cr}
Bel, L. 1958, \bibinfo{title}{CR Acad. Sci. 247 1094 Bel L 1962,} Cah. de Phys, 16, 59

\bibitem[{W.~B. Campbell \& T. Morgan(1971)Campbell \& Morgan}]{campbell1971debye}
Campbell, W.~B., \& Morgan, T. 1971, \bibinfo{title}{Debye potentials for the gravitational field,} Physica, 53, 264

\bibitem[{C. Caprini {et~al.}(2009)Caprini, Durrer, \& Servant}]{caprini2009stochastic}
Caprini, C., Durrer, R., \& Servant, G. 2009, \bibinfo{title}{The stochastic gravitational wave background from turbulence and magnetic fields generated by a first-order phase transition,} Journal of Cosmology and Astroparticle Physics, 2009, 024

\bibitem[{B. Carter(1985)Carter}]{carter1985canonical}
Carter, B. 1985, \bibinfo{title}{The canonical treatment of heat conduction and superfluidity in relativistic hydrodynamics.,} A Random Walk in Relativity and Cosmology, 48

\bibitem[{F. Cattorini \& B. Giacomazzo(2024)Cattorini \& Giacomazzo}]{cattorini2024grmhd}
Cattorini, F., \& Giacomazzo, B. 2024, \bibinfo{title}{GRMHD study of accreting massive black hole binaries in astrophysical environment: A review,} Astroparticle Physics, 154, 102892

\bibitem[{I. Ciufolini \& J.~A. Wheeler(1995)Ciufolini \& Wheeler}]{ciufolini1995gravitation}
Ciufolini, I., \& Wheeler, J.~A. 1995, Gravitation and inertia (Princeton university press)

\bibitem[{L.~F.~O. Costa \& J. Nat{\'a}rio(2021)Costa \& Nat{\'a}rio}]{costa2021frame}
Costa, L. F.~O., \& Nat{\'a}rio, J. 2021, \bibinfo{title}{Frame-dragging: meaning, myths, and misconceptions,} Universe, 7, 388

\bibitem[{C. Di~Castro \& R. Raimondi(2004)Di~Castro \& Raimondi}]{di2004disordered}
Di~Castro, C., \& Raimondi, R. 2004, \bibinfo{title}{Disordered electron systems,} in The Electron Liquid Paradigm in Condensed Matter Physics (IOS Press), 259--333

\bibitem[{A.~D. Dolgov {et~al.}(2002)Dolgov, Grasso, \& Nicolis}]{dolgov2002relic}
Dolgov, A.~D., Grasso, D., \& Nicolis, A. 2002, \bibinfo{title}{Relic backgrounds of gravitational waves from cosmic turbulence,} Physical Review D, 66, 103505

\bibitem[{A. Einstein(1915)Einstein}]{einstein1915erklarung}
Einstein, A. 1915, \bibinfo{title}{Erkl{\"a}rung der Perihelbewegung des Merkur aus der allgemeinen Relativit{\"a}tstheorie,} Sitzungsberichte der preu{\ss}ischen Akademie der Wissenschaften, 831, 839

\bibitem[{R. Fern{\'a}ndez {et~al.}(2019)Fern{\'a}ndez, Tchekhovskoy, Quataert, Foucart, \& Kasen}]{fernandez2019long}
Fern{\'a}ndez, R., Tchekhovskoy, A., Quataert, E., Foucart, F., \& Kasen, D. 2019, \bibinfo{title}{Long-term GRMHD simulations of neutron star merger accretion discs: implications for electromagnetic counterparts,} Monthly Notices of the Royal Astronomical Society, 482, 3373

\bibitem[{S. Galtier(2016)Galtier}]{galtier2016introduction}
Galtier, S. 2016, Introduction to modern magnetohydrodynamics (Cambridge University Press)

\bibitem[{S. Galtier(2022)Galtier}]{galtier2022physics}
Galtier, S. 2022, Physics of Wave Turbulence (Cambridge University Press)

\bibitem[{S. Galtier {et~al.}(2020)Galtier, Laurie, \& Nazarenko}]{galtier2020plausible}
Galtier, S., Laurie, J., \& Nazarenko, S.~V. 2020, \bibinfo{title}{A plausible model of inflation driven by strong gravitational wave turbulence,} Universe, 6, 98

\bibitem[{S. Galtier \& S.~V. Nazarenko(2017)Galtier \& Nazarenko}]{galtier2017turbulence}
Galtier, S., \& Nazarenko, S.~V. 2017, \bibinfo{title}{Turbulence of weak gravitational waves in the early universe,} Physical review letters, 119, 221101

\bibitem[{A. Garcia-Perciante {et~al.}(2009)Garcia-Perciante, Garcia-Colin, \& Sandoval-Villalbazo}]{garcia2009nature}
Garcia-Perciante, A., Garcia-Colin, L., \& Sandoval-Villalbazo, A. 2009, \bibinfo{title}{On the nature of the so-called generic instabilities in dissipative relativistic hydrodynamics,} General Relativity and Gravitation, 41, 1645

\bibitem[{M. Giovannini(2012)Giovannini}]{giovannini2012reynolds}
Giovannini, M. 2012, \bibinfo{title}{Reynolds numbers in the early Universe,} Physics Letters B, 711, 327

\bibitem[{G. Gogoberidze {et~al.}(2007)Gogoberidze, Kahniashvili, \& Kosowsky}]{gogoberidze2007spectrum}
Gogoberidze, G., Kahniashvili, T., \& Kosowsky, A. 2007, \bibinfo{title}{Spectrum of gravitational radiation from primordial turbulence,} Physical Review D, 76, 083002

\bibitem[{M. Hindmarsh {et~al.}(2021)Hindmarsh, L{\"u}ben, Lumma, \& Pauly}]{hindmarsh2021phase}
Hindmarsh, M., L{\"u}ben, M., Lumma, J., \& Pauly, M. 2021, \bibinfo{title}{Phase transitions in the early universe,} SciPost Physics Lecture Notes, 024

\bibitem[{W.~A. Hiscock \& L. Lindblom(1983)Hiscock \& Lindblom}]{hiscock1983stability}
Hiscock, W.~A., \& Lindblom, L. 1983, \bibinfo{title}{Stability and causality in dissipative relativistic fluids,} Annals of Physics, 151, 466

\bibitem[{P. Iroshnikov(1964)Iroshnikov}]{iroshnikov1964turbulence}
Iroshnikov, P. 1964, \bibinfo{title}{Turbulence of a conducting fluid in a strong magnetic field,} Soviet Astronomy, Vol. 7, p. 566, 7, 566

\bibitem[{T. Jacobson(1995)Jacobson}]{jacobson1995thermodynamics}
Jacobson, T. 1995, \bibinfo{title}{Thermodynamics of spacetime: the Einstein equation of state,} Physical Review Letters, 75, 1260

\bibitem[{A. Kosowsky {et~al.}(2002)Kosowsky, Mack, \& Kahniashvili}]{kosowsky2002gravitational}
Kosowsky, A., Mack, A., \& Kahniashvili, T. 2002, \bibinfo{title}{Gravitational radiation from cosmological turbulence,} Physical Review D, 66, 024030

\bibitem[{R.~H. Kraichnan(1965)Kraichnan}]{kraichnan1965inertial}
Kraichnan, R.~H. 1965, \bibinfo{title}{Inertial-range spectrum of hydromagnetic turbulence,} Physics of Fluids, 8, 1385

\bibitem[{L.~D. Landau \& E.~M. Lifshitz(1987)Landau \& Lifshitz}]{landau1987fluid}
Landau, L.~D., \& Lifshitz, E.~M. 1987, Fluid Mechanics: Volume 6, Vol.~6 (Elsevier)

\bibitem[{J. Liang {et~al.}(2024)Liang, Du, \& Xu}]{liang2024gravitomagnetohydrodynamicsspacetimeturbulenceearly}
Liang, J., Du, M., \& Xu, P. 2024, GravitoMagneto-Hydrodynamics and Spacetime Turbulence in Early Universe, \doarXiv{2404.14225}

\bibitem[{M. Liska {et~al.}(2020)Liska, Tchekhovskoy, \& Quataert}]{liska2020large}
Liska, M., Tchekhovskoy, A., \& Quataert, E. 2020, \bibinfo{title}{Large-scale poloidal magnetic field dynamo leads to powerful jets in GRMHD simulations of black hole accretion with toroidal field,} Monthly Notices of the Royal Astronomical Society, 494, 3656

\bibitem[{R. Maartens \& B.~A. Bassett(1998)Maartens \& Bassett}]{maartens1998gravito}
Maartens, R., \& Bassett, B.~A. 1998, \bibinfo{title}{Gravito-electromagnetism,} Classical and Quantum Gravity, 15, 705

\bibitem[{B. Mashhoon(2003)Mashhoon}]{mashhoon2003gravitoelectromagnetism}
Mashhoon, B. 2003, \bibinfo{title}{Gravitoelectromagnetism: a brief review,} arXiv preprint gr-qc/0311030

\bibitem[{C.~W. Misner {et~al.}(1973)Misner, Thorne, \& Wheeler}]{misner1973gravitation}
Misner, C.~W., Thorne, K.~S., \& Wheeler, J.~A. 1973, Gravitation (Macmillan)

\bibitem[{Y. Mizuno {et~al.}(2006)Mizuno, Nishikawaa, Koideb, Hardeec, \& Fishmand}]{mizuno2006grmhd}
Mizuno, Y., Nishikawaa, K., Koideb, S., Hardeec, P., \& Fishmand, G. 2006, \bibinfo{title}{GRMHD Simulations of Jet Formation with a Newly-Developed GRMHD Code,} in VI Microquasar Workshop: Microquasars and Beyond, 45--1

\bibitem[{M. Mo{\'s}cibrodzka {et~al.}(2009)Mo{\'s}cibrodzka, Gammie, Dolence, Shiokawa, \& Leung}]{moscibrodzka2009radiative}
Mo{\'s}cibrodzka, M., Gammie, C.~F., Dolence, J.~C., Shiokawa, H., \& Leung, P.~K. 2009, \bibinfo{title}{Radiative models of Sgr A* from GRMHD simulations,} The Astrophysical Journal, 706, 497

\bibitem[{R. Narayan {et~al.}(2012)Narayan, Sadowski, Penna, \& Kulkarni}]{narayan2012grmhd}
Narayan, R., Sadowski, A., Penna, R.~F., \& Kulkarni, A.~K. 2012, \bibinfo{title}{GRMHD simulations of magnetized advection-dominated accretion on a non-spinning black hole: role of outflows,} Monthly Notices of the Royal Astronomical Society, 426, 3241

\bibitem[{D.~A. Nichols {et~al.}(2012)Nichols, Zimmerman, Chen, Lovelace, Matthews, Owen, Zhang, \& Thorne}]{nichols2012visualizing}
Nichols, D.~A., Zimmerman, A., Chen, Y., {et~al.} 2012, \bibinfo{title}{Visualizing spacetime curvature via frame-drag vortexes and tidal tendexes. III. Quasinormal pulsations of Schwarzschild and Kerr black holes,} Physical Review D, 86, 104028

\bibitem[{D.~A. Nichols {et~al.}(2011)Nichols, Owen, Zhang, Zimmerman, Brink, Chen, Kaplan, Lovelace, Matthews, Scheel, {et~al.}}]{nichols2011visualizing}
Nichols, D.~A., Owen, R., Zhang, F., {et~al.} 2011, \bibinfo{title}{Visualizing spacetime curvature via frame-drag vortexes and tidal tendexes: General theory and weak-gravity applications,} Physical Review D, 84, 124014

\bibitem[{R. Penrose(1960)Penrose}]{penrose1960spinor}
Penrose, R. 1960, \bibinfo{title}{A spinor approach to general relativity,} Annals of Physics, 10, 171

\bibitem[{N. Peter {et~al.}(2018)Peter, Martin, \& G{\"u}nter}]{peter2018gravitational}
Peter, N., Martin, S., \& G{\"u}nter, S. 2018, \bibinfo{title}{Gravitational waves produced by compressible MHD turbulence from cosmological phase transitions,} Classical and Quantum Gravity, 35, 144001

\bibitem[{K.~S. Thorne {et~al.}(1986)Thorne, Thorne, Price, \& MacDonald}]{thorne1986black}
Thorne, K.~S., Thorne, K.~S., Price, R.~H., \& MacDonald, D.~A. 1986, Black holes: the membrane paradigm (Yale university press)

\bibitem[{C.~M. Will(1953)Will}]{will1953theory}
Will, C.~M. 1953, Theory and experiment in gravitational physics (Cambridge University Press (Cambridge, 1993))

\bibitem[{C.~M. Will(2014)Will}]{will2014confrontation}
Will, C.~M. 2014, \bibinfo{title}{The confrontation between general relativity and experiment,} Living reviews in relativity, 17, 1

\bibitem[{H. Yang {et~al.}(2015)Yang, Zimmerman, \& Lehner}]{yang2015turbulent}
Yang, H., Zimmerman, A., \& Lehner, L. 2015, \bibinfo{title}{Turbulent black holes,} Physical review letters, 114, 081101

\bibitem[{F. Zhang {et~al.}(2012)Zhang, Zimmerman, Nichols, Chen, Lovelace, Matthews, Owen, \& Thorne}]{zhang2012visualizing}
Zhang, F., Zimmerman, A., Nichols, D.~A., {et~al.} 2012, \bibinfo{title}{Visualizing spacetime curvature via frame-drag vortexes and tidal tendexes. II. Stationary black holes,} Physical Review D, 86, 084049

\end{thebibliography}
\bibliographystyle{aasjournalv7}

\end{document}